\documentclass[12pt]{article}
\usepackage{amsmath}
\usepackage{amssymb}
\usepackage{amsfonts}
\input amssym.def
\input amssym

\newcommand{\be}{\begin{equation}}
\newcommand{\ee}{\end{equation}}

\title{Electromagnetism and perfect fluids interplay in
multidimensional spacetimes}
\author{Nikolai V. Mitskievich\thanks{Physics Department, CUCEI,
University of Guadalajara, Guadalajara, Jalisco, Mexico.}
\thanks{Postal address: Apartado Postal 1-2011, C.P. 44100,
Guadalajara, Jalisco, M\'exico. E-mail:
mitskievich03@yahoo.com.mx}}
\date{~}
\begin{document}

\maketitle

\begin{abstract}
We consider fields in ($D>2$)-dimensional spacetime, whose
potential is $r$-form (skew-symmetric tensor of rank $r$), the
field tensor $F$ being its exterior derivative and the Lagrangian,
a function of the quadratic invariant $I$ of this tensor. It is
shown that vector field ($r=1$) describes electromagnetic field
only for $D=4$. In particular, for $D=3$ and the Lagrangean $L$ as
any function of the above-mentioned invariant, the ($r=1$)-field
has energy-momentum tensor identical with that of a perfect fluid
whose equation of state depends on the choice of $L(I)$.
\end{abstract}

Taking $D=n+1$ (see Ref. 3) and supposing $n$ to be an odd
integer, we find that there exist new natural ($r\neq
1$)-generalizations of the electromagnetic field. These
generalizations involve analogues of the dual conjugation of the
field tensor $F$ interchanging electric and magnetic parts of the
field, with its energy-momentum tensor invariant under this
conjugation. We consider such a field whose potential $A$ is
$r$-form, denoting this fields's invariant quadratic in the field
intensity $F$ (which has rank $r+1$) as $I$. For $L\sim I$, the
trace of the energy-momentum tensor identically vanishes (as this
was the case for $D=4$, the Maxwell field), so that the field is
intrinsically relativistic (see also my talk on the equivalence
principle in Session GT4 of MG11 where only $D=4$ was considered,
and Ref. 9). In these cases ${r=\displaystyle\frac{D}{2}}-1$. For
$D=4$ we have the usual electromagnetic field with its (not
intrinsically relativistic) nonlinear generalizations. $D=6$
yields $r=2$, and $F$ is skew-symmetric tensor of rank 3, {\it
etc}.

As to the multidimensional generalizations of perfect fluids, the
skew (axial = pseudo-) potentials have ranks $r=D-2$ and the field
tensors are their exterior derivatives \cite{Mits99a,Mits99b,
Mits03}. The equations of state are similar to those in $D=4$, and
the propagation of sound in multidimensional fluids is easily
described as behaviour of perturbations in these fluids closely
copying the situation in $D=4$. For example, there is the case of
$D$-dimensional stiff matter where sound propagates with the
velocity of light like in $D=4$. Moreover, in the special
relativistic limit it is easy to perform the second quantization
of small perturbations on the background of homogeneous fluid
(this was done in the unpublished Master of Sc. thesis of my
student H. Vargas Rodr{\'\i}guez in 1998) yielding phonons. This
occurs not only in $D=4$ case, but also in general, in particular
in $D=3$ which shows that the vector field describes there perfect
fluids and not electromagnetism (see final comments in this talk).

The stress-energy tensor of an $r$-form field follows from the
Noether theorem \cite{Mits58,Mits06}. In general it takes the form
\be \label{Tmunu} \mathfrak{T}^\beta_\alpha:=
\frac{\delta\mathfrak{L}}{\delta g_{\mu\nu}}g_{\mu\nu}
|^\beta_\alpha \equiv \frac{\delta\mathfrak{L}}{\delta g^{\mu\nu}}
g^{\mu\nu}|^\beta_\alpha=\frac{\delta\mathfrak{L}
}{\delta\left(|g|^{\frac{1}{2(r+1)}}g^{\mu\nu}
\right)}\left.\left(|g|^{\frac{1}{2(r+1)}}
g^{\mu\nu}\right)\right|^\beta_\alpha, \ee since the Lagrangian
density, as well as the function $L= \mathfrak{L}/\sqrt{|g|}\sim
I^k$, depend on $g^{\mu\nu}$ only algebraically (the $r$-form
potentials are considered to be independent of the metric tensor),
and $\left.\left(|g|^{\frac{1}{2(r+1)}} g^{\mu\nu}\right)
\right|^\beta_\alpha$ is the Trautman coefficient \cite{Traut}.
Then the intrinsically relativistic property condition
$T^\alpha_\alpha=0$ yields \be \label{intrel}
\left.\left(|g|^{\frac{1}{2(r+1)}}g^{\mu\nu}\right)
\right|^\alpha_\alpha=|g|^{\frac{1}{2(r+1)}}
g^{\mu\nu}\left(2-\frac{D}{k(r+1)}\right)=0 ~  ~  \Rightarrow ~  ~
k=\frac{D}{2(r+1)}. \ee When $k=1$, only space-times of even
number of dimensions $D$ can fit this condition: $D=2(r+1)$. The
same condition determines the conformal invariance property of the
fields. Thus in the intrinsically relativistic case it is
necessary and sufficient to use the simplest nonlinear Lagrangian
densities (see the Table \ref{table1}), \be \label{2.3.1}
\mathfrak{L}= \sqrt{|g|}\sigma I^k, ~ ~ ~ k=\frac{D}{2(r+1)}. \ee

\begin{table}       
\caption{Values of $k$ versus $r$ and $D$ describing intrinsically
relativistic fields.}
{\begin{tabular}{@{}l|ccccccccccc@{}} \hline\\
&&&&&&&&&&&\\[-18pt]
~~$r\backslash D$ & 2 & 3 & 4 & 5 & 6 & 7~ & 8 & 9~ & 10~~ & 11~ & 12~~ \\
\hline \hline ~~0 & {\bf 1} & 3/2 & 2 & 5/2 & 3 & 7/2~ & 4 & 9/2~
& 5 & 11/2~ & 6~ \\ \hline ~~1 & 1/2 & 3/4 & {\bf 1} & 5/4 & 3/2 &
7/4~ & 2 & 9/4~ & 5/2 & 11/4~ & 3~ \\
\hline ~~2 & ~ & 1/2 & 2/3 & 5/6 & {\bf 1} &
7/6~ & 4/3 & 3/2~ & 5/3 & 11/6~ & 2~ \\
\hline ~~3 & ~ & ~ & 1/2 & 5/8 & 3/4 & 7/8~ & {\bf 1} & 9/8~ & 5/4
& 11/8~ & 3/2~ \\
\hline ~~4 & ~ & ~ & ~ & 1/2 & 3/5 & 7/10 & 4/5 & 9/10 & {\bf 1}
& 11/10 & 6/5~ \\
\hline ~~5 & ~ & ~ & ~ & ~ & 1/2 & 7/12 & 2/3 & 3/4~ & 5/6 & 11/12
& {\bf 1}~ \\
\hline ~~6 & ~ & ~ & ~ & ~ & ~ & 1/2~ & 4/7 & 9/14 & 5/7 & 11/14 & 6/7~ \\
\hline ~~7 & ~ & ~ & ~ & ~ & ~ & ~ & 1/2 & 9/16 & 5/8 & 11/16 & 3/4~ \\
\hline ~~8 & ~ & ~ & ~ & ~ & ~ & ~ & ~ & 1/2~ & 5/9 & 11/18 & 2/3~ \\
\hline ~~9 & ~ & ~ & ~ & ~ & ~ & ~ & ~ & ~ & 1/2 & 11/20 & 3/5~ \\
\hline ~10 & ~ & ~ & ~ & ~ & ~ & ~ & ~ & ~ & ~ & 1/2~ & 6/11 \\
\hline ~11 & ~ & ~ & ~ & ~ & ~ & ~ & ~ & ~ & ~ & ~ & 1/2~ \\
\hline
\end{tabular}   \label{table1}}
\end{table}

\noindent This Table simply gives values of $k$; since $0\leq
r\leq D-1$, the lower left corner consists of blank spaces only:
the ``missing'' $r$-form field potentials are either trivially
exact ones, or equal to zero. $D=2$ is introduced here, of course,
only formally since the left-hand (``geometric'') side of
Einstein's equations then identically vanishes.

In particular, these results yield a\\
{\bf Theorem}: {\it (Generalized) fundamental electromagnetic
fields exist only in even-$D$-dimensional spacetimes, then being
$(r=D/2-1)$-form fields (see boldface {\bf 1}'s in the Table).
They possess all essential properties of the 3+1 Maxwell fields
(are linear, intrinsically relativistic, conformally invariant,
and subject to the $D$-dimensional dual conjugation relations).}

However, like in $D=4$, one has to introduce a new fundamental
field (alongside with the electromagnetic one) for any $D$ in
order to describe rotating perfect fluids (as well as
incompressible matter like that filling the interior Schwarzschild
solution). We call it the Machian field since it also determines
the cosmological term which vanishes in the intrinsically
relativistic case (phantom field). The Machian field also has a
skew potential, now with $r=D-1$. As we see in the Table
\ref{table1} above, its $k$ in the intrinsically relativistic case
is equal to $1/2$.

Thus we consider two fields (Machian, $r=D-1$, and Maxwellian,
$r=D/2-1$) as {\it fundamental} physical fields. All other
$r$-fields in the Table seem to be of less general importance; for
example, the ($r=D-2$)-field models perfect fluids in the
respective spacetimes, and its Lagrangian needs to be chosen as
such a function of the field invariant which yields the desired
equation of state (thus the relativistic property does not have a
deep meaning in the purely {\it phenomenological} case of fluids).
A plausible conjecture then is that both fundamental $r$-form
fields are bound to be intrinsically relativistic (the {\it free}
Machian field then yields the cosmological constant $\Lambda=0$
\cite{Mits99a,Mits99b}). A peculiar property of perfect fluids in
any $D$ is that in their field theoretical description the
inhomogeneity term in the corresponding field equations has the
sense of rotation and not of a source as this is the case for
electromagnetic and gravitational fields. This fact mathematically
follows from the definition of rotation as $u\wedge du$, the polar
covector $u=\ast F/I^{1/2}$ being four (or $D$-) velocity of the
fluid. Precisely this fact is completely overlooked in
``standard'' consideration of ($r=1$)-fields in $D=3$ where they
are erroneously treated as ``electromagnetic'' ones \cite{BHTZ,
Carlip,Garcia,KamKoi} (to mention only few of a vast number of
publications).

\end{document}